\documentclass[aps,showpacs,twocolumn,prb]{revtex4}
\usepackage{bm}
\usepackage[T1]{fontenc}
\usepackage{graphicx}
\usepackage{times}
\usepackage{amssymb,latexsym}
\begin{document}
\title{Metal-insulator transition in two-dimensional disordered systems with
power-law transfer terms}
\author{H. Potempa} \author{L. Schweitzer}
\affiliation{Physikalisch-Technische Bundesanstalt, Bundesallee 100, 38106
Braunschweig, Germany}
\date{\today}

\begin{abstract}
We investigate a disordered two-dimensional lattice model for
noninteracting electrons with long-range power-law transfer terms
and apply the method of level statistics for the calculation of the
critical properties. The eigenvalues used are obtained numerically by
direct diagonalization. We find a metal-insulator transition for a system
with orthogonal symmetry. The exponent governing the divergence 
of the correlation length at the  transition is extracted
from a finite size scaling analysis and found to be 
$\nu=2.6\pm 0.15$.
The critical eigenstates are also analyzed and the distribution of the
generalized multifractal dimensions is extrapolated. 
\end{abstract}
\maketitle

Starting with Anderson's work \cite{And58} more than four decades ago, 
the disorder driven metal-insulator transition carries on to be a
subject of active research. The underlying basic concept has proven to
be important in many physical situations because it provides
an essential requirement for the microscopical understanding of certain
quantum phase transitions in condensed matter physics. According to 
the scaling theory of localization,\cite{AALR79} noninteracting electrons 
in infinite one-dimensional (1D) and two-dimensional (2D) disordered
systems are always localized at zero temperature. 
It is known that there are exceptions to this rule. In particular, 
the quantum Hall system (absence of time reversal symmetry) with its 
diverging localization length in the center of the Landau bands 
\cite{Bodo} 
and the 2D disordered system with symplectic symmetry 
\cite{Sympl,SZ97} 
are notable examples.
However, even in the presence of time reversal symmetry, indications of a 
metal-insulator transition have been reported  
in experiments on two-dimensional electron and hole systems.\cite{AKS00} 
These findings have further intensified the efforts to 
look for possible mechanisms that allow a metal-insulator transition 
in systems with spatial dimension $d<3$. 
Clearly, electron-electron interactions that
were not taken into account in Ref.~2 
are important at low 
temperatures and possibly may account for the observed behavior.\cite{AKS00} 

Another class of systems exhibiting a metal-insulator transition for
noninteracting particles in dimensions $d < 3$, can be represented by models 
that in contrast to the Anderson model include long-range transfer terms. 
This possibility was already indicated in Anderson's paper,\cite{And58} 
but later addressed only a few times. For instance, 
the decay of the localized eigenstates has been investigated
numerically for a 1D tight-binding system with long-range transfer 
terms.\cite{YO87} Also, the localization
and dephasing of dipolar excitons in topologically disordered systems 
and the influence of weak long-range hopping on the 3D Anderson model 
have been studied \cite{LW87,Lev89,Lev90} as well as the
problem of Bloch states for a particle moving fast through a lattice
of Coulomb scatterers with power-law singularity.\cite{AL97} 
Instead of triggering the transition by varying the disorder strength, 
the crossover from localized to extended states has to be tuned by
changing the exponent $\beta$ of the power law transfer terms 
$V_{ij}\sim |\bm{r}_i-\bm{r}_j|^{-\beta}$.\cite{FM95,KM97a,Lev99}

Meanwhile, it has become clear that the behavior at the critical point 
and its multifractal eigenfunction statistics may be described by 
a one-parameter random matrix ensemble.\cite{KM97a} 
An intensively studied example is the 
power-law random banded matrix model (PRBM),\cite{MFDQS96,Kra99} 
which exhibits a transition from localized to extended states and 
contains multifractal eigenstates at the critical point. 
This was demonstrated by mapping the PRBM onto a nonlinear sigma model,
which could be solved for limiting cases.  In one thereof,
the PRBM transforms to the 1D version of the model 
studied by Levitov.\cite{Lev89,Lev90} The corresponding critical 
properties have been investigated numerically recently.\cite{CGO01}

Despite this considerable success, up to now it was not possible to
solve the experimentally more important 2D models with long-range 
transfer terms analytically. Therefore, no quantitative estimates have
been obtained for the critical quantities like the exponent $\nu$ of 
the correlation length or the fractal correlation dimension $D(2)$. 
The latter essentially influences the 
dynamics at the metal-insulator transition.\cite{Eta} 
In this paper we present the first results for the multifractal properties 
and the critical exponent of a 2D system with orthogonal symmetry, 
which was found to be $\nu=2.6\pm 0.15$.

The model investigated here describes noninteracting particles on a 
2D square lattice with distance dependent random transfer terms
$V_{ij}=V\varepsilon_{ij}\,(b/|\bm{r}_i-\bm{r}_j|)^{\beta}$ and 
random on-site potentials $\varepsilon_i$. 
The corresponding Hamilton operator is given by
\begin{equation}
{\cal H}=\sum_i\varepsilon_i|\bm{r}_i\rangle\langle \bm{r}_i|+
\sum_{i\ne j} V_{ij} 
|\bm{r}_i\rangle\langle \bm{r}_j|.
\label{Model}
\end{equation}
The $\{\varepsilon_i\}$ and $\{\varepsilon_{ij}\}$ are two sets 
of uncorrelated random numbers uniformly distributed between $-W/2$ and
$W/2$, and between $-S/2$ and $S/2$, respectively.
We take $b=1$, $S/V=1$, and fix the diagonal disorder by $W/V=6$. 
The unit of energy is set by $V=1$. 
The special choice of the diagonal disorder strength $W$ places the 
system in a certain volume of the parameter space where
the transition is easier to access.
The transfer terms, which allow transitions from a given lattice site 
to all others, depend on the corresponding distance via the power-law 
$|\bm{r}_i-\bm{r}_j|^{-\beta}$. By changing $\beta$ one tunes the
metal-insulator transition.
$L$ is the size of the square system measured in lattice spacings $a$, 
and Dirichlet boundary conditions are applied in both directions.

To investigate the localization behavior we apply the level statistics
method which has proven to be a very powerful tool in the 
past.\cite{Sea93,LevStat} 
The required eigenvalues for the level statistics and the eigenvectors
for the multifractal analysis were calculated by direct diagonalization. 
For the above chosen ratio of the diagonal and nondiagonal disorder, 
the density of states is symmetrical about the band center $E/V=0$ 
and only weakly energy dependent within the range [$-2V, 2V$]. 
The number of realizations calculated was such that for each set of 
parameters the total number of eigenvalues on the average add up to 
$6\cdot 10^5$ for system size $L/a\le 40$ and $3\cdot 10^5$ for larger
sizes. 
As in previous work, we choose the quantity $I_0(\beta,L)$ as our 
scaling variable because this choice does not depend on any arbitrary 
cutoff parameter.\cite{ZK95b,SZ97}
Here, $I_0(\beta,L)=\langle s^2\rangle$/2 is half the
second moment, $\langle s^2\rangle=\int_{0}^{\infty}s^2P_{\beta,L}ds$, 
of the probability density $P_{\beta,L}(s)$ for finding an energy 
difference of two consecutive eigenvalues, $s=|E_n-E_{n+1}|/\Delta$, 
where $\Delta$ denotes the mean level spacing.
We performed a proper unfolding procedure and checked that the results
were independent of the width of the energy interval around $E/V=0$
from which the eigenvalues were taken.

The eigenvalue statistics has been calculated within the energy interval
[$-1.5V, 1.5V$] for square systems of linear size $L/a=10, 15, 20, 25, 30,
40, 45, 50,55, 60,70, 80, 100$, and $120$.
The exponent $\beta$ of the power-law decay of the transfer terms 
was varied in the range $1.7\le\beta\le 2.3$. 
A total of 284 $I_0(\beta,L)$ values have been accumulated.
Increasing the size $L$, the magnitude of $I_0$ decreases for $\beta < 2$, 
but gets larger for $\beta > 2$ with an almost scale independent 
value at the point of intersection near $\beta = 2$. 
The latter behavior is a signature of a critical level statistics
connected with a quantum critical point \cite{Sea93}. 
In the limit $L\to\infty$, the scaling variable $I_0$ can take on
three values. For $\beta < 2$ only extended states are expected so 
that $I_0 = 0.643$ for orthogonal symmetry that is the 
universal random matrix result for the diffusive regime. For $\beta > 2$ 
all states will be localized in the thermodynamic limit so that $I_0 = 1.0$
which corresponds to the Poisson probability density distribution,
$P(s)=\exp(-s)$. This means that the probability of having two
neighboring eigenvalues close together is maximal, which is due to the
negligible overlap of the associated localized eigenvectors. The third
possible value $I_0$ can assume in the limit $L\to\infty$ is the
critical value $I_0^c$ at $\beta_c=2$, which is nonuniversal and known
to depend on model parameters as well as on the boundary conditions and  
on the shape of the  system.\cite{Boundary} 

\begin{figure}
\centering
\includegraphics[width=8.5cm]{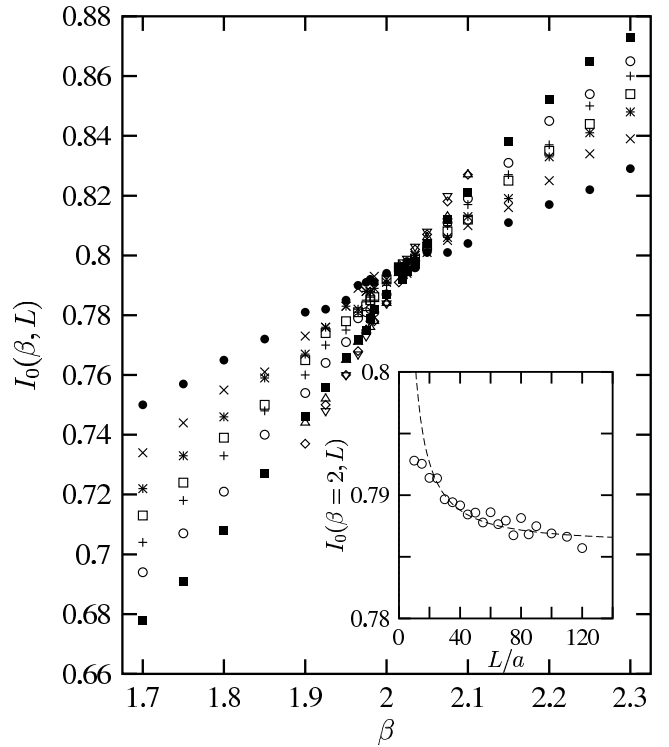}
\caption[]{$I_0(\beta,L)$ of the level statistics for a 2D disordered 
system with power-law transfer terms. The system size $L/a$ is
$10\ (\bullet)$, $15\ (\times)$, $20$\ ({\large$\ast$}), $25\ (\Box)$, 
$30\ (+)$, $40$\ ({\large$\circ$}), 
$60\ ({\scriptstyle \blacksquare})$ for power-law
exponents in the range $1.7\le\beta\le 2.3$, and $L/a=80\
(\vartriangle)$, $100\ (\Diamond)$ with $1.9\le\beta\le 2.1$,
and $120\ (\triangledown)$ with $1.925\le\beta\le 2.075$. 
The inset shows the size dependence of the $I_0(\beta=2,L)$ data which
can be fitted by the relation $I_0(\beta=2,L)=I_0^c+cL^{y}$ 
with $I_0^c\simeq 0.78$ and $y\simeq -1.2$.} 
\label{2d_O_raw}
\end{figure}

\begin{figure}
\centering
\includegraphics[width=8.5cm]{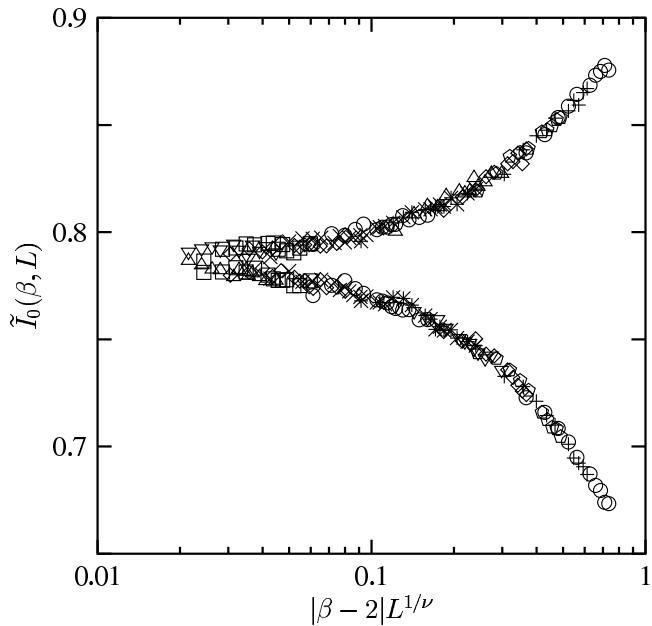} 
\caption[]{Scaling of the variance $\widetilde{I}_0(\beta,L)$ of the level
statistics for a 2D disordered system with power-law transfer terms.
The system size $L/a$ varies between $10$ and $120$ and 
the power-law exponent is taken from the range $1.7\le\beta\le 2.3$. 
The finite size scaling analysis yields 
$I_0^c=0.784 \pm 0.005$, a critical exponent $\nu=2.6\pm 0.15$, 
and an irrelevant scaling exponent $y=-1.25\pm 0.25$.}
\label{I0_2Dscal}
\end{figure}

\begin{figure}
\centering
\includegraphics[width=8.5cm]{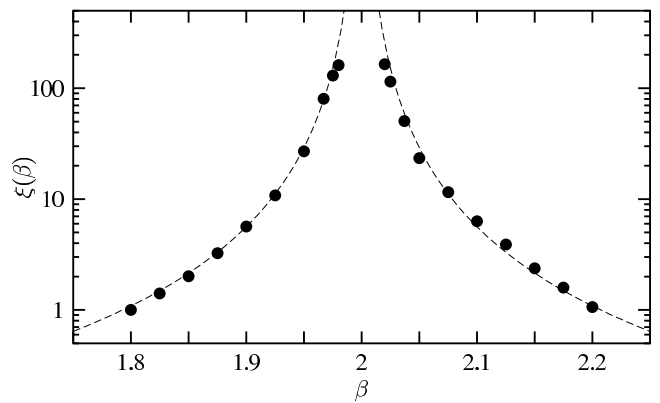} 
\caption[]{The dependence of the correlation length $\xi(\beta)$ on 
the power-law exponent $\beta$. The values are, up to a prefactor, 
calculated directly from the $\widetilde{I}_0(\beta,L)$ data which 
are the raw data $I_0(\beta,L)$ adjusted for irrelevant scaling
terms. The dashed line shows the best fit $\xi(\beta)\sim|\beta-2|^{-2.4}$.} 
\label{scal_MKBatsch}
\end{figure}

In Fig.~\ref{2d_O_raw}, which for clarity shows only a subset of the 
raw data, a crossing of the $I_0(\beta,L)$ values belonging to different
system sizes $L$ can be recognized near $\beta=2$. 
However, the weak monotonous shift of the crossing point with increasing 
system size $L$ indicates an irrelevant scaling term. 
From the $I_0(\beta=2,L)$ data we get an estimate for the 
irrelevant scaling exponent $y$ (see inset of Fig.~\ref{2d_O_raw}) 
which then was finally determined together with the correlation
exponent $\nu$ using a renormalization group ansatz where the 
dimensionless quantity 
$I_0(\beta,L)$ is expressed as a function,\cite{Scal}
\begin{equation}
I_0(\beta,L)=f_1(A_1 L^{1/\nu})+A_2 L^y f_2(A_1 L^{1/\nu}),
\end{equation}
containing a relevant and the leading irrelevant scaling variable 
$A_1$ and $A_2$, respectively. 
The corresponding scaling exponents are $\nu$ and $y$.
The functions $f_1$, $f_2$ as well as the scaling variables are
expanded in Taylor series with $\gamma=(\beta-\beta_c)/\beta_c$,
\begin{equation}
I_0(\beta,L) \simeq \sum_{p=0}^2 (C_p + L^y D_p) (\gamma L^{1/\nu})^{p}
\label{Taylor}
\end{equation}
Here, $C_p$ and $D_p$ are the coefficients that control the relevant
and irrelevant scaling fields, where $C_0=I_0^c$ determines the critical
value of $I_0(\beta,L)$ at $\beta_c$. Assuming that the deviations between 
the data and the model are random and caused by uncorrelated 
statistical noise due to the inherent disorder and the finite number of
realizations, the best fit was attributed to the set of
parameters with minimal $\chi^2$ statistics.
Since the critical power-law exponent is known,\cite{Lev89,Lev90}
$\beta_c=d$, we can fix $\beta_c=2$ and obtain for the orthogonal 2D case 
$I_0^c=0.784 \pm 0.005$, and a critical correlation exponent 
$\nu=2.6 \pm 0.15$. 
The leading irrelevant scaling term is governed by an exponent
$y=-1.25 \pm 0.25$. The resulting scaling curve is shown in 
Fig.~\ref{I0_2Dscal} where the original data corrected for the irrelevant
scaling terms,
$\widetilde{I}_0(\beta,L)=
I_0(\beta,L)-L^y\sum_{p=0}^2D_p(\gamma L^{1/\nu})^p$,
are plotted versus $|\beta-2|L^{1/\nu}$.
We achieve almost the same results if only 178 data points from the 
restricted interval $1.8\le \beta \le 2.2$ are taken into account. In
this case the quadratic terms in Eq.~(\ref{Taylor}) can be neglected. 
To test whether our ansatz for the correlation length,
$\xi\sim |\beta-\beta_c|^{-\mu}$, is justified at all, we directly
calculated $\xi(\beta)$ from our adjusted data using the method
described in Ref.~24. 
The result, which is shown in Fig.~\ref{scal_MKBatsch}, 
is consistent with our assumption and the best fit to the calculated data
yields $\mu=2.4\pm 0.1$. It is well known that with this method \cite{MK83}
one gets an estimate only for the lower bound of the true critical
exponent $\nu$. It is important to mention that our result for the
correlation length $\xi(\beta)$ in 2D differs completely from the 
asymmetrical function proposed for the corresponding 1D 
model.\cite{MFDQS96,CGO01}

Our final result for the critical exponent, $\nu=2.6\pm 0.15$,
is larger than the value accepted for the quantum Hall system  
$\nu=2.35\pm 0.03$,\cite{Bodo} but lies within the range of those 
reported for systems with symplectic symmetry (see, e.g., Ref.~25).
Also, we would like to remark that our result for the correlation
exponent is in good agreement with the value $z\nu=2.6\pm 0.8$
obtained recently from temperature scaling in an experiment on a 2D 
system where the disorder was generated by a layer of self-assembled 
quantum dots.\cite{Rea99} Thereby, one has to assume $z=1$ for the 
dynamical exponent as expected for interacting electrons. 
However, there are in the literature also reports of smaller
\cite{Kea96} as well as larger values \cite{Sea98a} for the
correlation exponent $\nu$ in 2D electron and hole gases.
Provided that the accordance between the experimental value \cite{Rea99} 
and our result for the model presented above is not fortuitous, 
the coincidence indicates that both metal-insulator transitions 
belong to the same universality class. 

\begin{figure}[b]
\centering
\includegraphics[width=8.6cm]{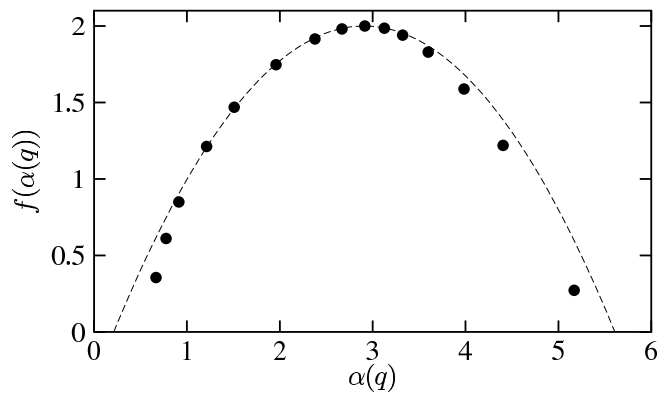} 
\caption[]{The $f(\alpha(q))$-distribution of a critical eigenstate
with energy at the band center for a square system of linear size
$L/a=150$. The parabolic approximation
$f(\alpha)=d-(\alpha-\alpha(0))^2/(4(\alpha(0)-d))$ 
is fitted by a single parameter $\alpha(0)=2.92\pm 0.05$.}
\label{falfa_2D}
\end{figure}

Now, having determined the critical exponent of the correlation
length for our model with power-law transfer terms, we turn to 
the analysis of the critical 
eigenstates $(\beta=2)$ for a square system of linear size $L/a=150$ 
and eigenenergy near zero (band center). 
A multifractal analysis has been performed along the standard procedure
where the scaling of a ``box-probability'' is calculated, 
$P(q,\lambda)=\sum_i^{N(l)}(\sum_{r\in\Omega_i(l)}|\psi_E(r)|^2)^q
\sim \lambda^{\tau(q)}$, from which the
generalized fractal dimensions $D(q)=\tau(q)/(q-1)$ or, 
by a Legendre transform, the so called $f(\alpha(q))$-distri\-bution  
can be derived.\cite{Fract} 
Here, $\Omega_i(l)$ is the $i$th box of 
size $l=\lambda L$ from which the $q$th moment of the
modulus of the normalized eigenstate $\psi_E(r)$ is taken.
In Fig.~\ref{falfa_2D} we show the $f(\alpha(q))$ distribution 
in comparison with the parabolic approximation, 
$f(\alpha(q))=d-(\alpha(q)-\alpha(0))^2/(4(\alpha(0)-d))$,
which for finite systems is valid only for small $|q|$.\cite{Parabol} 
As usual in studies on finite systems, 
the deviations are larger for such $\alpha(q)$ that correspond to 
negative $q$ values because the negative exponents blow up those 
spots where the eigenstate almost vanishes. 
The correlation dimension is extracted to be $D(2)=0.9\pm 0.05$ 
so that the exponent $\eta=d-D(2)\simeq 1.1$ 
describing anomalous diffusion near the critical point is very large
compared to the other known 2D situations, e.g., $\eta\simeq 0.38$
for the quantum Hall system \cite{Eta} 
and $\eta\simeq 0.35$ for the symplectic case.\cite{Sch95} 
Therefore, one has to expect a strong influence on the dynamical
properties at the metal-insulator transition originating from the 
spatial amplitude fluctuations of the critical eigenfunctions.\cite{BHS96}

In conclusion, we investigated a disordered two-dimen\-sional 
lattice model with long-range transfer terms which are governed
by a power-law decay $\sim |\bm{r}_i-\bm{r}_j|^{-\beta}$. 
Using the method of level statistics, the 
existence of a metal-insulator transition has been demonstrated in
a 2D system with orthogonal symmetry. 
The transition is tuned by the power-law exponent $\beta $ of the 
transfer terms. From a finite size scaling analysis the
critical exponent determining the divergence of the correlation 
length near the transition has been obtained. 
We found $\nu=2.6\pm 0.15$ which is very close 
(assuming a dynamical exponent $z=1$) to an experimental result,
$z\nu=2.6\pm 0.8$, which was estimated recently from temperature 
scaling.\cite{Rea99}

We thank Ferdinand Evers for very helpful discussions.


\end{document}